\documentclass{cargese}\usepackage{graphicx}
\let\footnote\savefootnote
\let\footnotetext\savefootnotetext 
\let\footnoterule\savefootnoterule 
\normallatexbib
\def\bea{\begin{eqnarray}} 
\def\eea{\end{eqnarray}}
\def\beann{\begin{eqnarray*}} 
\def\eeann{\end{eqnarray*}}
\def\beq{\begin{equation}} 
\def\eeq{\end{equation}}
\def\ba{\begin{array}} 
\def\ea{\end{array}}
\def\ben{\begin{enumerate}} 
\def\een{\end{enumerate}}
\def\4{\tilde }
\def\5{\bar }  
\def\6{\partial } 
\def\7{\hat } 
\def\ep{\epsilon}
\def\g{\gamma} \def\G{\Gamma}
\def\m{\mu}
\def\n{\nu}
\font\mybb=msbm10 at 10pt
\def\bb#1{\hbox{\mybb#1}}

\def\bE {\bb{E}}

\def\bM {\bb{M}}

\begin{document}
%------------ article title  ------------------->>

% For a long title use \\ to cut lines.
% In that case, supply  alternate version of the title
% in square brackets, (it will go in the Table of contents during final
% production of the book.
% \articletitle[Short title]{The long version \\ of this title}

\articletitle[]{Supersymmetric Worldvolume \\ Solitons}

%% optional, to supply a shorter version of the title for the running head:
%%\chaptitlerunninghead{}

\author{Joan Sim\'{o}n}

%% multiple authors at the same institution may be separated with \\
%% like in \author{Samuel Bostaph\\
%%                 and Gregor Kariotis}

\affil{Departament ECM, Facultat de F\'{\i}sica \\
Universitat de Barcelona and Institut d'Altes Energies \\
Diagonal 647, E-08028 Barcelona, Spain}         
% Your Institution and address. May cut into  separated 
%                 % lines with \\
%
% optional email address
\email{jsimon@ecm.ub.es}
%
%%% Repeat the above for multiple authors at different institutions.
%% \author{ }
%% \affil{ }
%% \email{ }

% optional abstract
\begin{abstract}
The general criteria for finding bosonic supersymmetric worldvolume
solitons is reviewed. We concentrate on D-branes, discussing in
particular, bion/dyon solutions and D3 branes on NS5 backgrounds.
\end{abstract}

%------------ body of article ------------------->>
% Write your article here. 
% Note that the \section{section title}
% command allows for the form \section[short title]{very long\\ title}
% Idem for \subsection and \subsubsection
%------------ end of article ------------------->>

\paragraph{Introduction}
\index{introduction}

Brane effective actions constitute a relativistic extension of supersymmetric
field theories, in the sense that the energy density functional ${\cal E}$
satisfies ${\cal E}^2 = f(\phi^i)$, which upon linearisation matches the
conventional susy free field theory result. This observation suggests that
all supersymmetric field theories might be realized in terms of branes
and their intersections. It is desirable to establish a complete classification
of their classical solutions, especially the BPS ones that will receive no
quantum corrections but contribute to the path integral.

A universal feature of all these effective actions is the existence of a
fermionic gauge symmetry called kappa symmetry. The latter is not only
responsible of removing half of the fermionic degrees of freedom, but it
also restricts the background geometry to be compatible with the supergravity
equations of motion and relates spacetime supersymmetry with worldvolume
supersymmetry. In particular, any bosonic supersymmetric configuration
should satisfy \cite{BOP} :
\beq
\G_\kappa\,\ep = \ep
\label{e1}
\eeq
where $\G_\kappa$ is the matrix appearing in $\kappa$-symmetry transformations,
satisfying $\G_\kappa\,^2=1$ and Tr$\,\G_\kappa=0$, while $\ep$ is an arbitrary
linear combination of Killing spinors of the background. The number of
supersymmetries preserved by the combined background/brane system is the 
number of linearly independent solutions of (\ref{e1}). Notice that if
the background admits no Killing spinors, the worldvolume theory will never
be supersymmetric.

It is important to stress that (\ref{e1}) is equivalent to the calibration
approach \cite{GPT} for those branes with no excited gauge field living on
their worldvolumes. This is the reason why we will concentrate on Dp-branes
whose $\G_\kappa$ matrix can be written as a $p+1$-form (d-form) :
\[
\sqrt{-det\,({\cal G}_{\m\n} + {\cal F}_{\m\n})}\,\G_\kappa = 
e^{\cal F}\wedge\,\bigoplus_{n=0}^{(2d-1+(-)^d)/4}   
\,\g_{2n + {1+(-1)^p\over 2}}{\cal P}_i\,^{n +{1+(-1)^p\over 2}}
\label{kappa0}
\]
where $\g_\m=\6_\m X^m e_m^a\G_a$ while ${\cal P}_A = \G_{11}$ and
$({\cal P}_B)^s = (\sigma_3)^s i\sigma_2$.

Condition (\ref{e1}) depends on the background $(e^a_m,\, B_{mn},\,
\ep)$ and on the configuration $(X^m,\, A_\m)$. Once the background
is fixed, its solution will not only determine the number of independent
components of $\ep$ but will provide us with an equation(s), that will play
the role of BPS equation(s), relating the different excited fields. Once the
BPS equation(s) is known, one can always write the square of the energy
density as a sum of positive definite terms that allow us to find some
bound on the energy \cite{GGT}. It is precisely when the BPS equation(s)
is satisfied that the bound is saturated. Since the energy is minimised,
it is no longer necessary to check the equations of motion, but the Gauss'
law and/or the Bianchi identity, that will determine the equation of motion 
for the remaining independent excited fields.

\paragraph{Bions}
\index{bions}

Let us consider Dp-branes probing Dp-branes geometry $(1\leq p \leq 7)$. The
background is given by 

\bea
ds^2 & = &  U^{-1/2}\,ds^2(\bM_{(1,p)}) +  
U^{-1/2}\,ds^2(\bE_{9-p})\label{metric} \\
e^{\phi} & = & U^{(3-p)/4} \quad \, \quad C_{0\ldots p} = U^{-1}
\label{dp}
\eea
where the gauge choice for the R-R potential is consistent with the no force
condition existing between parallel Dp-branes and $U$ is an harmonic
function in the transverse space to the branes.

To describe a F-string ending on a probe parallel to a stack of Dp-branes
$(X^\m = \xi^\m \,,\,\m=0,\ldots ,p)$, we must excite a transverse scalar
$(Y=Y(\xi^a)\,,\, a=1,\ldots ,p)$ and the electric components of the gauge
field $A_0=A_0(\xi^a)$, since the F-string is an electrically charged
NS-NS object. It is natural to demmand $\G_{0\ldots p}
{\cal P}_i^{(2d+1-(-)^d)/4}\ep_0 =  \ep_0$ and $\G_{0y}\G_{11}\ep_0 = -\ep_0$
for type IIA and $\G_{0y}\sigma_3 \ep_0 =\ep_0$ for type IIB, 
on the assymptotic value of the Killing spinor $(\ep_0)$, since these
are the conditions describing the existence of a Dp-brane and a F-string,
respectively. 

When eq. (\ref{e1}) is subject to the above constraints, it is solved
iff
\beq
F_{0a} = \6_a Y \quad \Leftrightarrow \quad \4E^a = \delta^{ab}\6_b Y
\label{bion}
\eeq
where $\4E^a$ is the conjugate momentum of the gauge field. (\ref{bion})
is the BPS equation as it can be checked by analysing the energy density
\cite{BT} of the configuration 
\beq
({\cal E} + U^{-1})^2= (U^{-1}+\4E^a\6_a Y)^2 + U^{-1}|\4E^a - \delta^{ab}
\6_b Y|^2\, .
\eeq 
The Gauss' law $(\6_a \4E^a=0)$ determines the eq. of motion $(\delta^{ab}
\6_a\6_b Y=0)$ for the excited scalar. Its solution and the corresponding
physical interpretation depend on the worldvolume dimension.

Due to the self-duality of $G_{(5)}=d\,C_{(4)}$, one should expect dyonic
solutions in the case of D3-branes. The soliton condition on $\ep_0$ is
modified to $\G_{0y}(\cos\alpha\,\sigma_3 + \sin\alpha\,\sigma_1)\ep_0 = 
\ep_0$, while (\ref{e1}) is solved iff $F_{0a}=\cos\alpha\,\6_a Y$ and 
$B^a=\sin\alpha\,\delta^{ab}\6_b Y$. The excitation of the magnetic field 
is consistent with the charge carried by the D-strings.

The latter analysis could be extended to D8-branes, since there is no new
contribution due to the mass parameter for the configurations under study.

\paragraph{D3-branes in NS5 backgrounds}
\index{NS5back}

Let us consider a D3-brane probing NS5-branes geometry. The background is
given by :
\bea
ds^2 & = & ds^2(\bM_{1,5}) + U\,\left(dr^2 + r^2\,d\Omega_3^2\right) \\
e^{2\phi} & = & U  \quad , \quad H_{(3)} = 2R^2\,\omega_{(3)}
\label{back}
\eea

We will describe the S-dual configuration to the one considered in 
\cite{GRST} (to which we refer for notation), but from the D3-brane 
perspective. The D3-brane "wraps" the transverse $S^3$ sphere and
the NS5-branes magnetic charge induces a magnetic charge on the probe
through a D-string soliton connecting both. In this way, the assymptotic
Killing spinor $\ep_0$ should satisfy $\G_{0\ldots 5}\,\sigma_3 \ep_0 = 
\ep_0$, $\G_{0\theta\theta_2\theta_3}\,i\sigma_2 \ep_0 = \ep_0$ 
and $\G_{0r}\,\sigma_1 \ep_0 = -\ep_0$, that correspond to the background,
probe and soliton susy projection conditions.

Due to the polar coordinates on $\bE^4$, $\ep$ is not constant. Proceeding
as in \cite{GRST}, eq. (\ref{e1}) is solved if :
\bea
B^{\7a} & = & -\Delta {\5g^{\7a\7b}\6_{\7b}r \over r} \nonumber \\
B^{\theta} & = & {-\Delta \over (r\sin\theta)'}\left(
(r\cos\theta)' - \5g^{\7a\7b}{\6_{\7a}r\6_{\7b}(r\sin\theta)\over r}
\right)
\label{sol}
\eea
where $\Delta=U\,r^2\,\sqrt{det\,\5g}$ and $B^a={1\over 2}\ep^{abc}
{\cal F}_{bc}$. Eqs. (\ref{sol}) are the BPS equations as can be checked
by writting the energy density as a sum of squares :
\bea
{\cal E}^2  & = & |r\,B^{\7a} + \Delta\,\5g^{\7a\7b}\6_{\7b}r|^2 +
\left(B^a\6_a (r\sin\theta) + \Delta(r\cos\theta)'\right)^2 + {\cal Z}^2 \\
{\cal Z} & = & \Delta(r\sin\theta)' - B^a\6_a (r\cos\theta)
\eea

The Bianchi identity $d{\cal F} = -d\7B$, which is equivalent to
$\6_a B^a=2R^2\,\,\sqrt{det\,\5g}$, not only gives the corresponding equation
of motion for the excited radial scalar, but it is also useful to check
the topological character of ${\cal Z}$ which can be written as
${\cal Z} = \6_a {\cal Z}^a$, where 
\bea
{\cal Z}^{\7a} & = & -B^{\7a}r\cos\theta \\
{\cal Z}^{\theta} & = & -B^\theta r\cos\theta + \sqrt{det\,\5g}\sin\theta
\left({r^3\over 3} + rR^2\right)
\eea

A similar analysis to the one carried in \cite{GRST} shows that for
$SO(2)$ invariant configurations, the above soliton gives a worldvolume
realization of the Hannany-Witten effect.

\begin{acknowledgments}
J. S. would like to thank the organizers for a very nice school, and
for financial support. This work was supported by a fellowship from
Comissionat per a Universitats i Recerca de la Generalitat de Catalunya
and by AEN98-0431 (CICYT), GC 1998SGR (CIRIT).
\end{acknowledgments}

%% appendix optional
%\chapappendix{This is the Appendix Title}
%This is an appendix with a title.

%\chapappendix{}
%This is an appendix without a title.

%

\end{document}